\pdfoutput=1
\documentclass{article}

\usepackage[final]{neurips_2019}

\usepackage[utf8]{inputenc} 
\usepackage[T1]{fontenc}    
\usepackage{hyperref}       
\usepackage{url}            
\usepackage{booktabs}       
\usepackage{amsfonts}       
\usepackage{nicefrac}       
\usepackage{microtype}      
\usepackage{graphicx}       
\usepackage{caption}        
\usepackage{subcaption}     
\usepackage{floatrow}
\usepackage{bm}
\usepackage{pdfpages}

\makeatletter
\newcommand{\printfnsymbol}[1]{%
  \textsuperscript{\@fnsymbol{#1}}%
}
\newfloatcommand{capbtabbox}{table}[][\FBwidth]

\makeatother

\title{Interactive Image Restoration}

\author{%
  Zhiwei Han, Thomas Weber, Stefan Matthes \thanks{These authors contributed equally to this research.}\\
  fortiss \\
  Guerickestraße 25, 80805 München \\
  \texttt{\{han, weber, matthes\}@fortiss.org} \\
   \And
   Yuanting Liu, Hao Shen \\
   fortiss \\
   Guerickestraße 25, 80805 München \\
  \texttt{\{liu, shen\}@fortiss.org} \\
}
\begin{document}

\maketitle

\begin{abstract}
Machine learning and many of its applications are considered 
hard to approach due to their complexity and lack of transparency.
One mission of human-centric machine learning is to improve 
algorithm transparency and user satisfaction while ensuring an
acceptable task accuracy. In this work, we present an interactive 
image restoration framework, which exploits both image prior and 
human painting knowledge in an iterative manner such that they can 
boost on each other. Additionally, in this system users can 
repeatedly get feedback of their interactions from the restoration 
progress. This informs the users about their impact on the 
restoration results, which leads to better sense of control, which can lead to greater trust and approachability. 
The positive results of both objective 
and subjective evaluation indicate that, our interactive approach 
positively contributes to the approachability of restoration 
algorithms in terms of algorithm performance and user experience.
\end{abstract}

\section{Introduction}
\begin{figure}[h]
    \begin{minipage}{.24\textwidth}
        \centering
        \includegraphics[width=\textwidth]{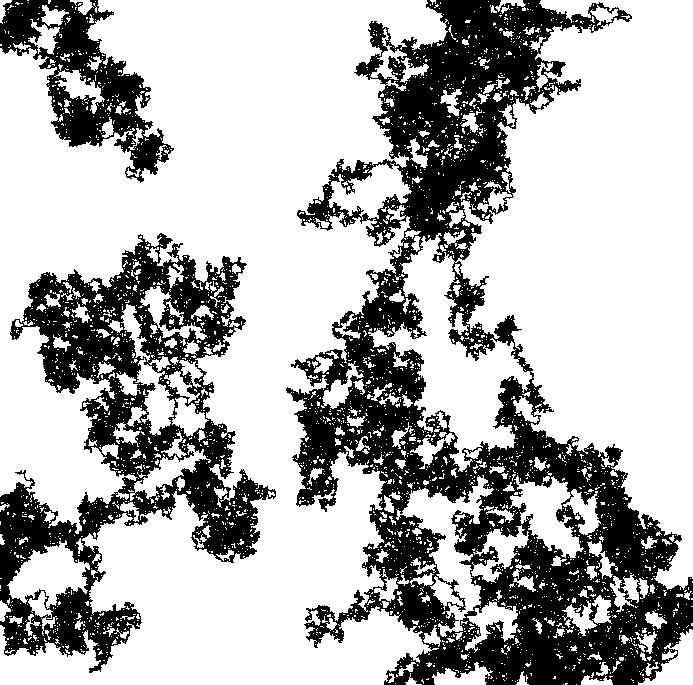}
    \end{minipage}%
    \begin{minipage}{.24\textwidth}
        \centering
            \includegraphics[width=\textwidth]{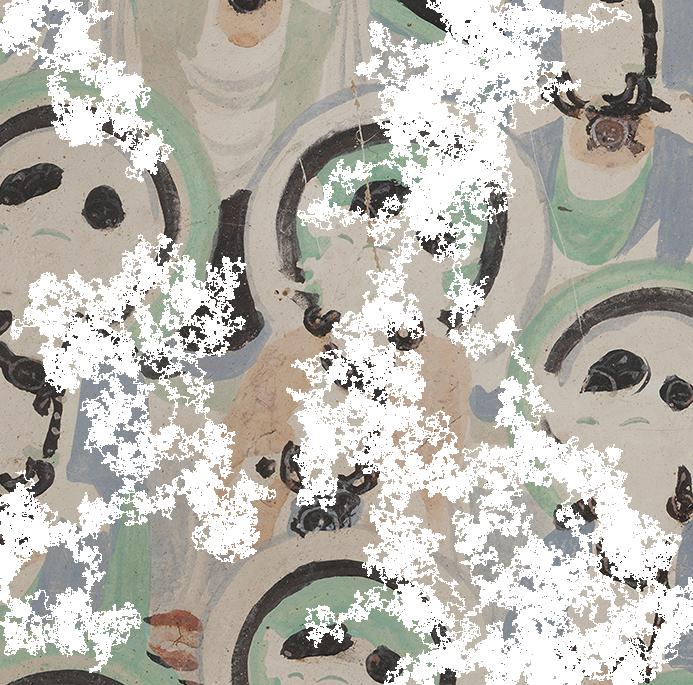}
    \end{minipage}
    \begin{minipage}{.24\textwidth}
        \centering
            \includegraphics[width=\textwidth]{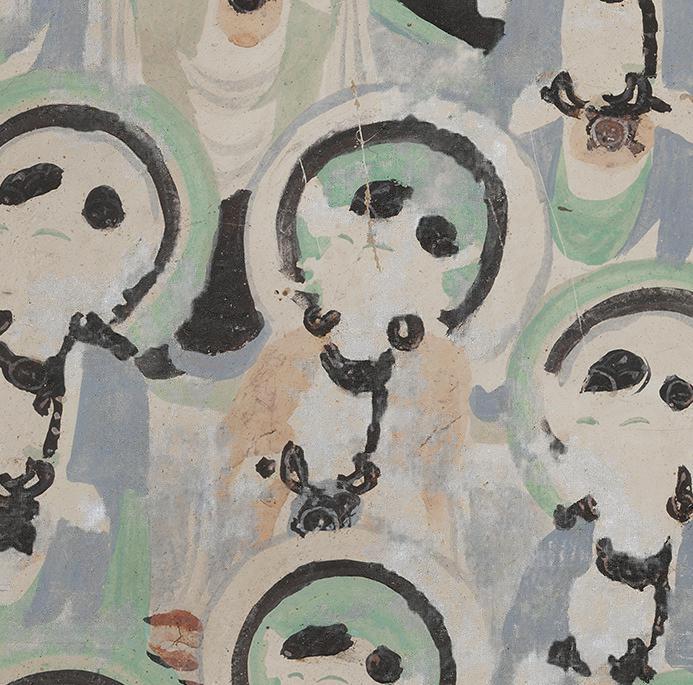}
    \end{minipage}
    \begin{minipage}{.24\textwidth}
        \centering
            \includegraphics[width=\textwidth]{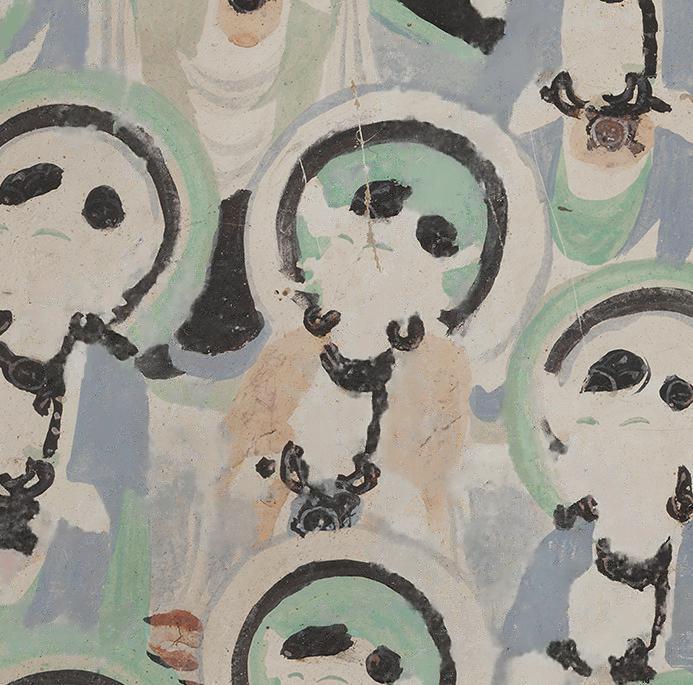}
    \end{minipage}
    \caption{Left to right: An artificially generated damage mask, 
    the image damaged within the masked area, the image restored by Deep Image Prior \cite{ulyanov2018deep} and the restored image using our interactive approach.}
    \label{image_demo}
\end{figure}

Image inpainting is a process for restoring damaged or missing sections of images, 
such that the results are visually plausible. Naturally, 
performance of restoration algorithm degrades when the corrupted 
sections become dense or large, since more semantic information is 
missing. 

Due to the lack of semantic information, restored images can contain artifacts 
like areas with inconsistent texture or monotone color as shown in the third 
image of Fig. \ref{image_demo}. Despite this, given those pre-restored images, human beings can easily deduce the semantics 
in the corrupted regions. Therefore, this awareness can be used to 
accomplish restoration tasks. Based on this intuition, we extend Deep 
Image Prior (DIP) \cite{ulyanov2018deep} with Human-Computer 
Interaction (HCI) and present Interactive Deep Image Prior (iDIP), 
a collaborative, interactive image restoration framework 
(Sect. \ref{approach}). This framework enables human and algorithms 
to collaboratively restore images in an iterative manner. With the 
proposed framework, even people with little painting knowledge can 
generate plausible images and manage restoration task. 
Furthermore, frequent feedback promises higher sense of control 
and better user satisfaction than non-interactive methods.

We then evaluate iDIP-based image restoration system with respect to two 
research questions:
\begin{enumerate}
  \item Does the interactive approach produce higher quality images?
  \item How do users view such a system regarding user experience and satisfaction?
\end{enumerate}
We answer the first question in Sect. \ref{experiment} in terms of 
objective and subjective measurement. 
To judge user experience and satisfaction, we have conducted a user 
study as described in Sect. \ref{user_study}. 

\section{Related Work}
Previous research works attempted to fully automate
the image restoration process. As one of the state of 
the art approaches, DIP restores images by exploiting 
image prior modelled by a Convolution Neural Network 
(CNN) \cite{krizhevsky2012imagenet}. DIP minimizes the 
following loss function for image inpainting:
\begin{equation} \label{eqa}
    \mathcal{L} = \min_{\bm{\theta}} ||(f_{\bm{\theta}}(\bm{z}) - \bm{x}_0) \oslash \bm{m}_0||_2,
\end{equation}
where $f_{\bm{\theta}}$ is a CNN parameterized with $\bm{\theta}$, $\bm{z}$ 
is a fixed input, $\bm{x}_0$ is a corrupted image, $\oslash$ is Hadamard product
and $\bm{m}_0$ is the mask for damage area. DIP overcomes the drawbacks of 
exemplar-based \cite{barnes2009patchmatch, hays2007scene, kwatra2005texture, he2012statistics} 
and learning-based methods \cite{yu2018generative, iizuka2017globally, yeh2017semantic, yan2018shift}, 
such as difficulties in recovering sophisticated texture and 
requirement of large training set, respectively. 

Same as classic machine learning models, training of 
DIP is non-interactive and will be performed only once. 
However, DIP cannot use human 
understanding of textural semantics and leads to poor 
user satisfaction due to its low transparency. 
Nonetheless, interactive Machine Learning (iML) \cite{fails2003interactive} 
increases the sense of control by introducing human intervention
into learning loops \cite{amershi2014power}. The increased sense of control can improve trust and user experience in many scenarios 
\cite{amershi2014power, cohn2003semi, holzinger2016interactive, johnson2008active}.
\section{Approach}\label{approach}
To our best knowledge, there is no previous work 
combining DIP with iML. In this work, 
we extend the DIP with interactivity \cite{fails2003interactive} 
and bring humans into the training loops of iDIP. The 
updates of iDIP is iterative, focused and rapid. These 
properties make the restoration process more transparent 
and contribute to a user-satisfied approach (Sect.\ref{user_study}).

iDIP restores images by iteratively exploiting image prior 
and human knowledge via human-in-the-loop intervention. The underlying human involvement 
could be either creating new mask (correction) or painting on 
the corrupted regions (guidance).

\begin{figure}[h!]
  \begin{floatrow}
        \ffigbox{
        \includegraphics[width=.65\textwidth]{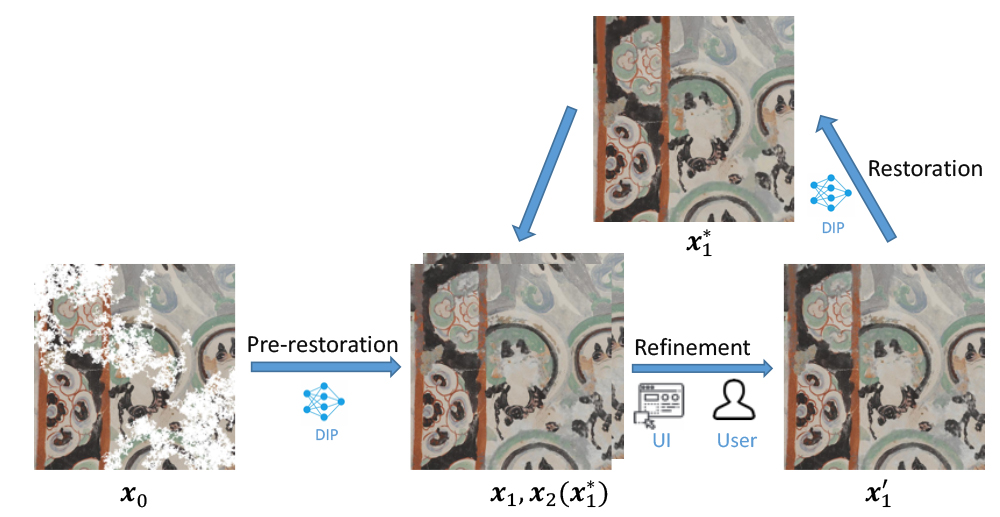}
    }{
        \caption{iDIP Training Iteration}
        \label{IDIPI} 
    }
    \hspace{1.5cm}
    \ffigbox{%
      \includegraphics[width=.34\textwidth]{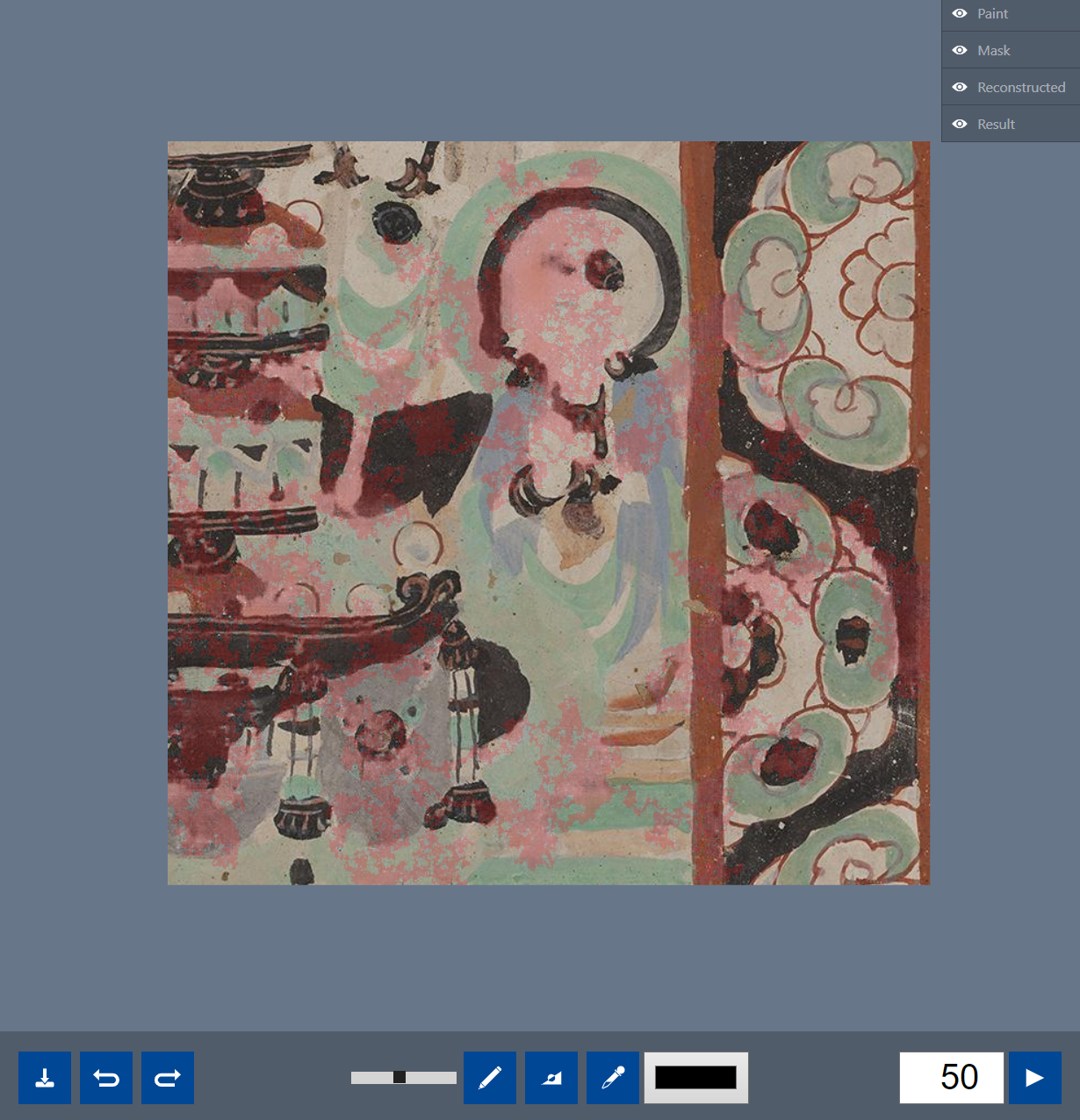}
    }
    { 
        \caption{User Interface}
        \label{UI}
        }
  \end{floatrow}
\end{figure}

\textbf{Training iteration}: One training iteration can be visualized 
in Fig. \ref{IDIPI} and it consists of three stages. 1. User is 
presented with the image $\bm{x}_n$ restored by iDIP from the last 
iteration, where $n$ is the current timestamp. 2. User paints on the 
image $\bm{x}_n$ to obtain a refined image $\bm{x}_n^{'}$. 3. iDIP 
restores image $\bm{x}_n^{'}$ by minimizing the loss function in Eq. 
\ref{eqa} and output the image $\bm{x}_n^*$. Note, the output image 
$\bm{x}_n^*$ of the $n$th iteration is equivalent to the input image 
$\bm{x}_{n+1}$ of the $(n+1)$th iteration.

Given pre-restored image $\bm{x}_n$, users can come up more easily 
with textural semantics in the damage region than only given $\bm{x}_0$. 
Furthermore, iDIP exploits its restoration performance by distilling 
the reconstructed textual information in the refined image $\bm{x}_n^{'}$. 
In this way, iDIP and human knowledge can jointly boost on each other.

Besides, this iterative approach endows users with better control of 
their impact through trial-and-error. Therefore, users can better 
determine their involvement intensity in next iterations. Frequent 
interaction contributes to better user satisfaction and system 
transparency. What's more, early stopping can be applied 
on time since users continuously observe the textural 
consistency and can terminate the process in any iteraction to avoid overfitting.

\textbf{User interface (UI)}: Fig. \ref{UI} shows the UI for iDIP. 
The image in center is the pre-restored one with the mask as red overlay. 
Users were supposed to pick appropriate color and paint in the masked region.

\section{Experiment}\label{experiment}
We conducted experiments to answer the first research 
questions in this section: \textit{Does the interactive 
approach produce higher quality images?}

\textit{\textbf{Dataset:}} We used the Dunhuang Grottoes Painting 
Dataset \cite{yu2019dunhuang} for the experiments. The dataset 
contains 500 full frame paintings with artificially generated masks 
for damage region, of which we randomly picked ten.

\textit{\textbf{Metrics:}}\label{experiment_metrics} As performance 
measurement, we computed Dissimilarity Structural Similarity 
Index Measure (DSSIM) \cite{wang2004image} and Local Mean 
Squared Error (LMSE) \cite{grosse2009ground} between restored and 
ground truth images. 
Mean Squared Error (MSE) and Structural Similarity Index Measure (SSIM) are common and easy-to-compute measures of the perceived quality of digital images or videos in computer vision. In this paper, we compute MSE and equalize it to LMSE by setting $k=1$. By using $DSSIM=\frac{1-SSIM}{2}$ we let the DSSIM also be inversely proportional to restoration quality as LMSE.

\textit{\textbf{Baselines:}} To show the effectiveness, we compared 
our approach with five state of the art baselines. For learning 
based methods, we used their pre-trained model on Places2 
\cite{zhou2017places}, because it is one of the widely-used 
scene recognition dataset.
\begin{itemize}
    \item \textbf{EdgeConnect}: EdgeConnect \cite{nazeri2019edgeconnect} 
    proposed a two-stage adversarial model and can deal with irregular 
    masks.
    \item \textbf{PartialConv}: PartialConv \cite{liu2018image} used 
    partial convolutions with an automatic mask update step.
    \item \textbf{PatchMatch}: PatchMatch \cite{barnes2009patchmatch} 
    can quickly find approximate nearest-neighbor matches 
    between image patches and was adopted by Photoshop.
    \item \textbf{PatchOffset}: PatchOffset \cite{he2012statistics} 
    minimizes an energy function to find patches with dominant offsets.
    \item \textbf{Deep Image Prior}: DIP \cite{ulyanov2018deep} 
    exploits the image prior by minimizing Mean Squared Error (MSE) in 
    the unmasked region.
\end{itemize}

For objective evaluation, we compared images restored by all six 
algorithms on ten randomly picked corrupted images using two metrics. The images generated by iDIP for the objective evaluation were recovered by domain expert. Each image was completed within 1200 iterations (600 iterations before painting and 600 after) and the domain expert painted only once per image. 
However, only pixel-wise 
measures can not account for human criteria used to judge the quality 
such as semantic correctness and consistency. In consequence we asked 
each of the 19 participants from our user study (described in the following section) to subjectively select two best 
recovered images out of the six produced by the six algorithms.

\subsection{Results}\label{experiment_results}
\begin{figure}
  \begin{floatrow}
    \capbtabbox{
    \begin{tabular}{l|ll}
      \toprule
      & DSSIM & LMSE \\
      \cmidrule(r){1-3}
      EdgeConnect & 0.2803 & 629.65 \\
      PartialConv & 0.2816 & 2550.02 \\
      PatchMatch & 0.2423 & $\textbf{185.68}^*$ \\
      PatchOffset & 0.2246 & 558.05 \\
      DIP & \textbf{0.2228} & 214.23 \\
      iDIP & $\textbf{0.2227}^*$ & \textbf{207.37} \\
      \bottomrule
    \end{tabular}    
    }{
      \caption{Comparison on restoration metrics}
      \label{result_table}
    }
    \ffigbox{%
      \includegraphics[width=0.5\textwidth]{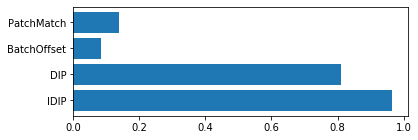}
    }
    { 
        \caption{Comparison of subjective opinions} 
        \label{fig_subj}
    }
  \end{floatrow}
\end{figure}

\textbf{Objective evaluation}: In the Tab. \ref{result_table} we 
can see that, although we initialized the networks with pre-trained 
weights, two learning-based methods still have the worst performance.
Style transfer failed because the image style of Dunhuang 
dataset varied too much from the training set. PatchMatch has 
the best LMSE score by a large margin. However, our approach 
slightly outperformed all non-interactive methods on DSSIM and 
has the second smallest LSME score. This suggests that 
interactivity positively contributes to output quality.

\textbf{Subjective evaluation}: Fig.\ref{fig_subj} shows the 
probability of one algorithm being picked as top two algorithm
in the subjective evaluation. We left out learning-based methods, since
they had not been picked. The two DIP-variants significantly outperformed other 
methods, even though PatchMatch demonstrated the best result on LMSE. 
Compared to DIP, iDIP still showed a considerable improvement, 
which indicates interactivity introduced in iDIP added to the
output quality. 

The difference between the two evaluations is also noteworthy:
While PatchMatch has the lowest LMSE score, subjectively it appears far inferior to the DIP-based methods. This may be an indicator that simple similarity measures are insufficient to account for human perception.

To summarize, introducing interactivity in iDIP positively affected the restoration performance. Therefore, we confidently give a positive answer to the first research question.
\section{User Study}\label{user_study}
With iDIP outperforming the other baselines in the subjective 
perception and being not far off with respect to objective 
measures, it remains whether an iML approach is attractive 
from a usage point of view. We evaluated this in a user study
and via a questionnaire.

Participants in this study (n = 19; 9 male, 9 female, 1 other; 20-29 years old: 10, 30-39 years old: 7, 40-49 years old: 2)
were people medium expertise with image 
manipulation (mean: 2.68/5, std: 1.25) and low expertise with image reconstruction (mean: 1.74/5, std: 1.19). We presented to them the UI and asked them to reconstruct two images. Due to practical reasons, we limited their working time to seven minutes per image. We then asked the participants to fill 
out the questionnaire regarding general satisfaction with the 
process using the System Usability Scale (SUS) and workload using 
NASA TLX as well as questions regarding the benefits of our 
interactive approach.
  
Results from the SUS and TLX were very positive (average score 
SUS: 86/100, TLX: 3.4/10). Measured on a 5-point Likert scale, 
the opinion of the participants regarding iML being suitable for 
image reconstruction (4.5/5) and in general (4.0/5) were also very 
positive. Participants also did not believe that a non-interactive
ML process (0.9/5) or a manual approach (1.8/5) would perform better.

The fact that all participants stated that they liked the 
combination of interactivity and machine learning, as well as other 
feedback, led us to conclude that iML can make machine learning more approachable. Whether it is an actual boost to expert-productivity  remains to be seen in future work.
\section{Conclusion and Future Work}\label{conclustion_future}
In this paper we have outlined our framework for interactive image
restoration. This framework allows users to interactively contribute to DIP-based 
image restoration process so that both image prior and human knowledge can be well leveraged in an iML fashion. Our experiments show that
the designed interactions positively affected the output quality 
as iDIP outperformed all five state of the art baselines. Meanwhile, good 
user satisfaction has been achieved according to the user study, as 
participants stated their appreciation and confidence of the proposed method.
In summary, the positive answers of two research questions indicate that 
our goal of human-centric machine learning have been fulfilled for image 
restoration tasks.

As human-in-the-loop approach demonstrated its effectiveness in terms of 
algorithm performance and user satisfaction, we remain the interpretation of 
rich interactions forms in image restoration as future work.


\bibliographystyle{unsrtnat}
\bibliography{references}
\newpage
\section{Supplementary Material}
As supplementary materials, we provide the subjective evaluation 
record of restoration performance, the questionnaire used in the 
user study and the statistical summary of user study.
\includepdf[pages=-]{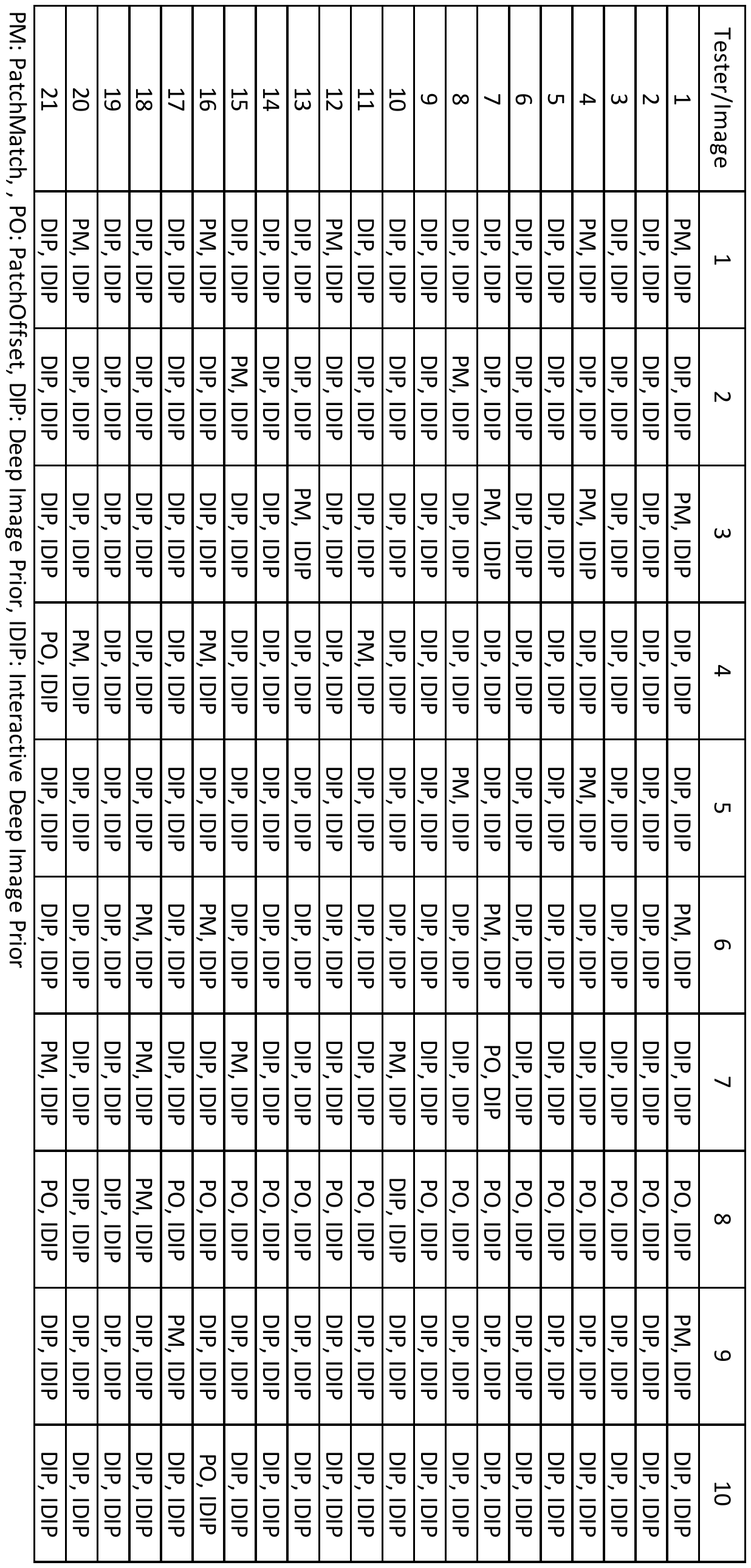}
\includepdf[pages=-]{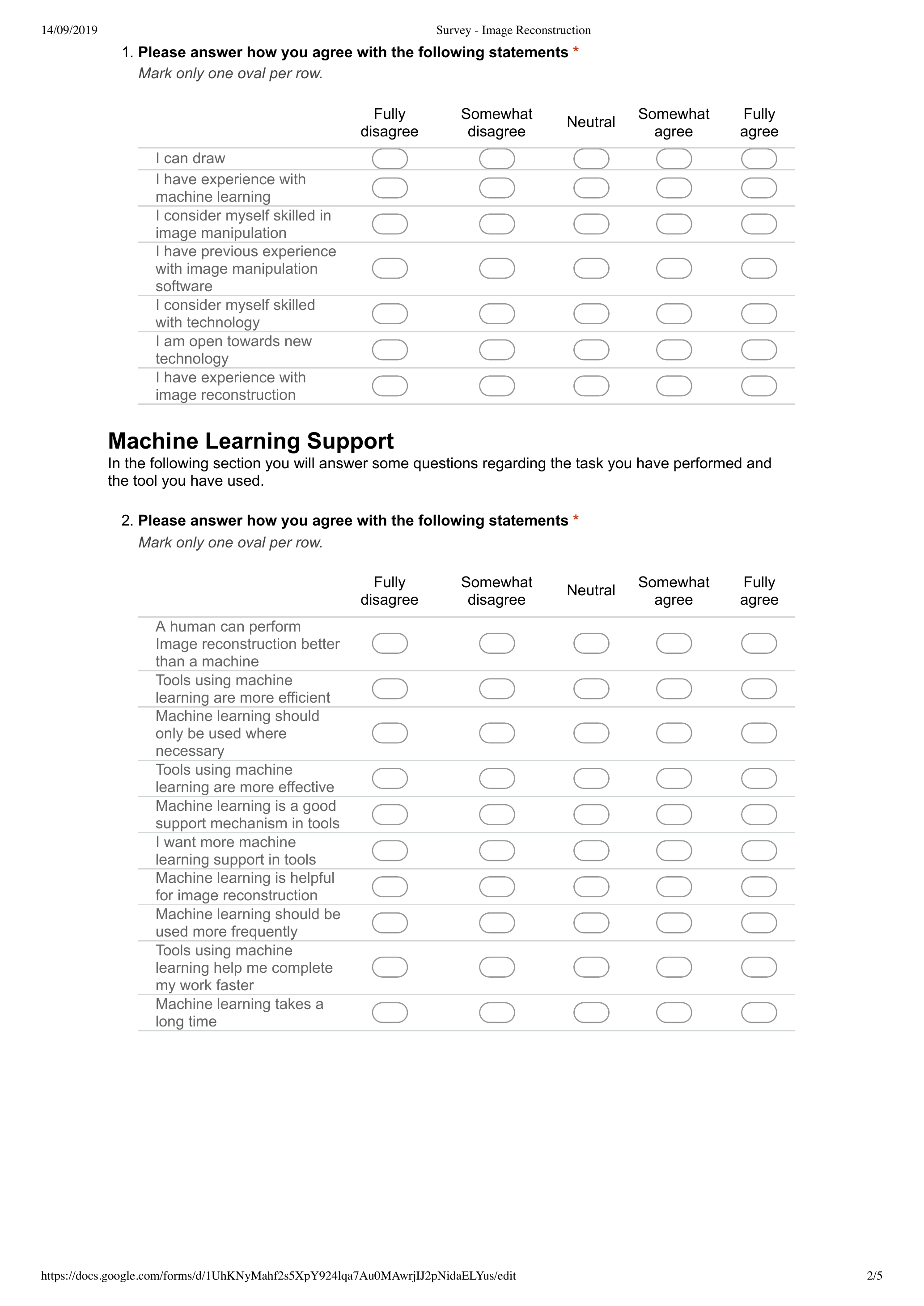}
\includepdf[pages=-]{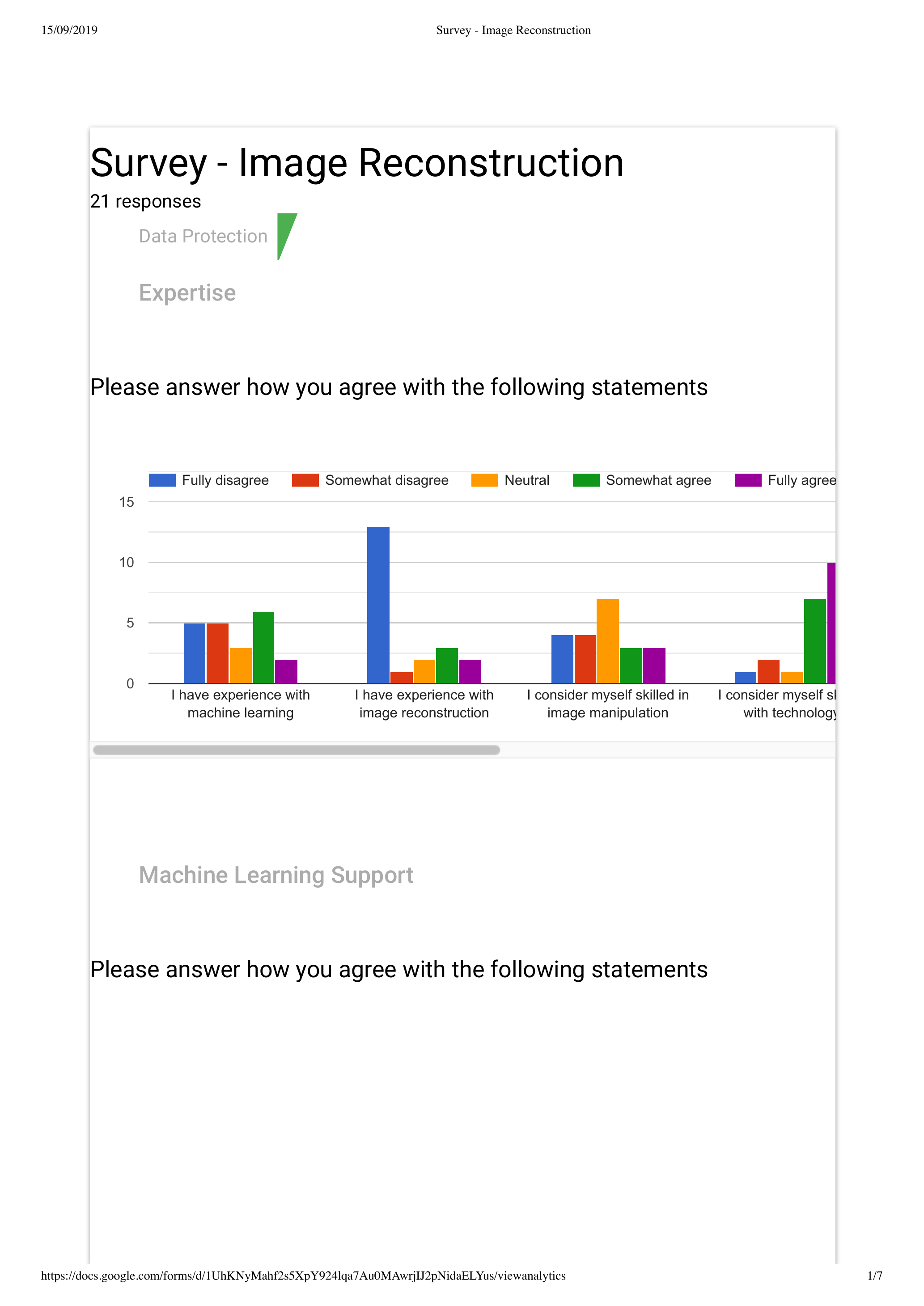}
\medskip

\small

\end{document}